# Mechanical Properties of End-crosslinked Entangled Polymer Networks using Sliplink Brownian Dynamics Simulations

## revised version

Julian Oberdisse[1,2], Giovanni Ianniruberto[2], Francesco Greco[3], Giuseppe Marrucci[2]

[1] Laboratoire des Colloïdes, Verres et Nanomatériaux (LCVN)
Université Montpellier II, F – 34095 Montpellier Cedex

[2] Dipartimento di Ingegneria Chimica
Università degli studi di Napoli "Federico II"

[3] Istituto per i Materiali Compositi e Biomedici - CNR
Piazzale Tecchio 80
I –80125 Napoli

**Tables:  3**
**Figures: 9**

**ABSTRACT**

The mechanical properties of a polymeric network containing both crosslinks and sliplinks (entanglements) are studied using a multi-chain Brownian dynamics simulation. We coarse-grain at the level of chain segments connecting consecutive nodes (cross- or sliplinks), with particular attention to the Gaussian statistics of the network. Affine displacement of nodes is not imposed: their displacement as well as sliding of monomers through sliplinks is governed by force balances.

The simulation results of stress in uniaxial extension and the full stress tensor in simple shear including the (non-zero) second normal stress difference are presented for monodisperse chains with up to 18 entanglements between two crosslinks. The cases of two different force laws of the subchains (Gaussian chains and chains with finite extensibility) for two different numbers of monomers in a subchain ($n_o = 50$ and $n_o = 100$) are examined. It is shown that the additivity assumption of slip- and crosslink contribution holds for sufficiently long chains with two or more entanglements, and that it can be used to construct the strain response of a network of infinitely long chains. An important consequence is that the contribution of sliplinks to the small-strain shear modulus is about ⅔ of the contribution of a crosslink.

## I. INTRODUCTION

The outstanding mechanical properties of polymer networks are due the entropic elasticity of long chain molecules linked together by permanent junctions called crosslinks (Treloar 1975). Theoretical understanding of rubber elasticity started with the affine and phantom network theories some 60 years ago (James and Guth 1947). These early theories conjecture affine displacement of the junctions, and evaluate its effect on otherwise unperturbed chains. Interactions between chains are neglected, i.e. these theories do not take entanglements into account. In the 1960's the tube model of rubber elasticity was set up by Edwards (1967). In this major contribution, entanglements where recognized to act as topological constraints on the chains. However, original tube theories are based on a single chain in a mean field (Doi and Edwards, 1986), the different types of which have been reviewed and generalized by Rubinstein and Panyukov (2002). A complete multi-chain network theory with crosslinks and entanglements modeled by sliplinks has been proposed by Ball et al (1981) for Gaussian chains, and later by Edwards and Vilgis (1988) for chains with finite extensibility.

Complex network topologies with both entanglements and crosslinks, randomly or end-crosslinked, are not easily described theoretically. A possible way out are computer simulations of bead-spring chains in the spirit of the work of Kremer and Grest (Kremer and Grest 1990, Duering et al 1994, Everaers and Kremer 1995). These simulations are very general because they are based only on the chain connectivity and on the excluded volume interaction between any monomer. Thus they do not use any hypothesis on the nature of entanglements themselves. The drawback of this method is that it is computationally expensive.

The purpose of this article is to provide a numerical implementation of the network theories by Ball et al (1981) and Edwards and Vilgis (1988). Unlike these underlying theories, our model can in principle be applied to any network structure, with an explicit description of entanglements. A major difference with atomistic simulations consists in the coarse-graining, which is done at the level of the subchains linking two entanglements in order to decrease the computational effort. Conceptually, this is similar to the successful 'Twentanglement' package, where entanglements are described on a scale (blobs) including many monomers (Padding and Briels 2001, Padding and Briels 2002, Padding and Briels 2003). Such a coarse-graining on the rheologically relevant length scale enables prediction of entangled melts in complex flows, e.g. shear flows (Padding and Briels 2003). One major difference between 'Twentanglements' and the work presented here is that we use the classical force law used in polymer science, whereas it is determined by microscopic simulations in the Twentanglement-model. Another difference between our and the atomistic (and Twentanglement) simulations is in the nature of the entanglement, which is represented as a strictly binary interaction by our sliplinks as well as in the theory of Ball et al (1981). In this respect, sliplinks are in fact closer to the "figure-of-eight" slide-ring connections in the recently synthesized polyrotaxane gels by Okumura and Ito (2001) and Karino et al (2005), though chemical crosslinks in the classical sense are absent in the latter case.

In a previous letter, we have presented first results obtained with our network model (Oberdisse et al 2002), a more elaborate version of which exists also for melts (Masubuchi et al 2001, Masubuchi et al 2003). It was found that the agreement with the underlying theories by Ball et al and Edwards and Vilgis is very good in the small deformation regime. Upon high deformation, however, our model predicted insufficient strain softening in uniaxial elongation. In the following, we find that a similar situation is also encountered in shear. These discrepancies between theory and simulation will be discussed in the final section.

In this article, we present a systematic study of the mechanical properties of monodisperse networks of various number of entanglements per chain. The predicted tensorial stress response of the system to deformation is reported in a parametrized form, necessitating no further numerical work. The article is organized as follows: In Section II we describe the construction of the network with entanglements, which are represented by sliplinks. The motion of the cross- and sliplinks is governed by a Langevin equation, as well as the sliding of chains through the slip-links. In Section III we report the statistical properties of the network at equilibrium, and the response to deformation in uniaxial extension and simple shear. Results are also discussed in terms of additivity of crosslink and sliplink contributions in Section IV.

## II. DESCRIPTION OF THE SIMULATION

### II.1 Network Construction

An amorphous network is constructed by creating $N_c$ monodisperse chains as random walks of steplength $\ell$ in a three dimensional cubic box, with periodic boundary conditions. Each step of a random walk represents a subchain connecting two beads, which is our level of coarse-graining of the real chain. To form a network, chains are then connected at the beads. The first and the last bead of a chain, the external beads, form crosslinks by connection to other external beads, whereas internal beads are connected to other internal ones to form sliplinks, cf. Fig. 1. These connections distort the chains, thereby altering their statistical properties. We therefore *bias* the random walk of the chains *before* connection, in order to obtain a real random walk *after* connection and equilibration (see below). The details of the algorithm used to bias the random walk are described in Appendix A. Note that the problem of equilibrating long chains by simulation is not specific to the present simulation (Auhl et al 2003).

We call Z the number of beads per chain, $N_s = Z - 1$ the number of subchains per chain, and the total number of subchains in the box is $N = N_c \cdot N_s$. The ratio $\nu = N/V$ (where $V = L_x L_y L_z$ is the box volume) is the density parameter. Similarly, one can define the number of beads per unit volume $\rho = N_c \cdot Z/V$, and write it as a sum of the density of internal and external beads $\rho = \rho_E + \rho_I$. The network is built up from the chains by connection of beads that are spatially close, where "closeness" is defined through appropriate search radii $R^E$ and $R^I$, different for external and internal beads. In order to assure equal chances of success of the connection attempts, we take:

$$R^E = 2 \, \rho_E^{-1/3} \tag{1a}$$

$$R^I = 2 \, \rho_I^{-1/3} \tag{1b}$$

Here we have arbitrarily chosen a prefactor of 2, because it leads to a reasonable number of possible partner beads in the search sphere. Alternatively, if only the closest possible connection was taken, a low final connectivity would result. Once a partner bead is found, beads are moved to a common position between the partners, and a node (slip- or crosslink) is formed. If all possible connections with the initial search radii are concluded, the search radii are increased by 20%, the search is continued, and so on until the desired functionality f is reached (f > 3.95). The general linking conditions are as follows: **(i)** the resulting crosslinks have a functionality f ≤ 4; **(ii)** a sliplink is made of a pair of internal beads (f = 4), and the two partner beads must not be nearest neighbours along the same chain; **(iii)** a network is accepted if pathologies are rare: less than 1% of unconnected chain ends, and less than 1.5% of chain ends in a crosslink with f = 2, which would amount to doubling the chain mass. The latter condition is readily fulfilled for short chains, while not so easily for long ones (Z ≥ 10). Once the network has been formed by connecting the chains, $n_o$ monomers are assigned to each subchain, and the system is equilibrated by switching on the dynamics. The resulting statistical properties of the network are tuned by carefully choosing the initial parameters (the steplength $\ell$, the bias parameter $\Theta_b$ and the density of beads $\rho$), as given in Appendix A. The final, equilibrated network has the following statistical properties:

$$<\mathbf{a}_i^2> = <n_i>b^2 = n_o\, b^2 \qquad (2a)$$

$$<\mathbf{R}_j^2> = N_s <\mathbf{a}_i^2> \qquad (2b)$$

where $\mathbf{R_j}$ and $\mathbf{a_i}$ are the end-to-end vectors of chains and subchains, respectively, b is the monomer (or Kuhn) length, and $n_i$ is the (fluctuating) number of monomers in a subchain, where $<n_i> = n_o$ by monomer conservation. The monomer length b defines the length scale and is set equal to 0.1 for convenience. Indeed, with $n_o$ set to 100, this gives an equilibrated average square subchain length of unity. Eqs. (2) express the fact that chains are Gaussian down to subchain level, as observed experimentally by Boué et al (1987).

## II.2 Dynamics

The dynamical equations are the same as in our letter (Oberdisse et al, 2002), and we recall them for convenience. Dynamics (Brownian dynamics) is simulated by *two types of motions* in our network of crosslinked chains with entanglements. The first one is the *motion of nodes* (slip- and cross-links), due to random (thermal) agitation and to the net force resulting from the subchains pulling on the node. The basis for the force is the classical linear spring law valid for Gaussian chains. In a first time, we will explore the properties of networks made up of chains governed by this linear force law. In a second part, we will take the finite extensibility of chains into account by introducing a non linear term $f(x_i)$ in the force, where $x_i = |\mathbf{x_i}|$ is the ratio of the end-to-end distance of a subchain and its contour length $n_i b$:

$$\mathbf{x}_i = \mathbf{a}_i / n_i\, b \qquad (3)$$

For chains with finite extensibility $x_i$ is smaller than (or at most equal to) one. For the $i^{th}$ subchain the force reads:

$$\mathbf{F}_i = \frac{3kT}{b} f(x_i)\, \mathbf{x}_i \qquad (4)$$

$$f(x) = \frac{1}{1 - x^2} \approx 1 + x^2 + x^4 + x^6 \qquad (5)$$

The nonlinear extension is derived by series expansion from the expression proposed by Warner (1972). The advantage of using the series expansion is that any overstretching of subchains which might possibly happen during a simulation run due to finite $\Delta t$ values does not lead to a numerical catastrophe (i.e. division by zero, or very high forces). In the case of a linear force law, f(x) reduces to one.

The discretised equation of motion in the Stokes limit then reads

$$\Delta \mathbf{r} = \frac{6 D \Delta t}{f \cdot b} \sum_{i=1}^{f} f(x_i) \mathbf{x_i} + \sqrt{\frac{12 D \Delta t}{f}} \mathbf{u} \tag{6}$$

where $\mathbf{r}$ is the node position, $\Delta t$ is the time step, f the node functionality, and $\mathbf{u}$ is a unit vector of random direction. The first term in eq. (6) arises from the subchains pulling on the node, and the second from the stochastic force. The direction of the latter is random, and we have adopted a fixed amplitude, taken in accordance with the fluctuation-dissipation theorem (Honerkamp 1994, Doi and Edwards 1986). In principle, the motion is governed by the diffusivity of the total node $D_n$, which we choose to express through the diffusivity D of an individual subchain: $D_n = 2$ D/f, i.e. f/2 subchains are supposed to contribute to the friction of the node. The Einstein equation D = kT/$\zeta$ relates the diffusivity to the friction coefficient $\zeta$ of one subchain, taken as a constant.

The second equation of motion describes the *sliding of chains through sliplinks*. The sliding of monomers results from a different tension in the two subchains belonging to the same chain entering and leaving an entanglement, as well as from a stochastic force. The sliding equation is derived from the one-dimensional motion of a virtual bead located at the entanglement position (s=0). This virtual bead is pulled to one side or the other ($\Delta s > 0$ or $< 0$), resulting in an exchange of monomers $\Delta n$. The discretised equations read

$$\Delta s = \frac{3 D \Delta t}{b} \left( x_j f(x_j) - x_i f(x_i) \right) \pm \sqrt{2 D \Delta t} \tag{7a}$$

$$\left. \begin{array}{ll} \Delta n_i = -n_i \dfrac{\Delta s}{\Delta s + a_i} & \text{if } \Delta s > 0 \\[2ex] \Delta n_j = -n_j \dfrac{-\Delta s}{-\Delta s + a_j} & \text{if } \Delta s < 0 \end{array} \right\} \tag{7b}$$

$$\Delta n_i + \Delta n_j = 0 \tag{7c}$$

where i and j indicate consecutive subchains along the same chain, and the one-dimensional random force is accounted for through the $\pm$ term. Note that the sliding process of eqs. (7) is described by a one-dimensional analog of eq. (6), with the same diffusivity of a subchain D. As only one subchain contributes to the friction in sliding, we have different numerical prefactors of the spring forces in eqs. (6) and (7). The stochastic term changes as well due to the different dimensionality. The equations for $\Delta n$ are written assuming constant monomer density along the subchain which *loses* monomers, and are such that n never becomes negative. Finally, eq. (7c) assures that the total number of monomers is a conserved quantity.

## II.3 Equilibration

Equilibration of the system is achieved by randomly picking a node and moving it according to eq. (6). If the node is a sliplink, then eqs. (7) are used first, once for each chain forming the entanglement. We checked that inverting the order of execution of eqs. (6) and (7) gives identical results. During these operations all other beads are kept fixed. Time is incremented by $\Delta t$ after all nodes have been moved once (on average). Approach to equilibrium is monitored through the total energy of the system. The system is run long enough for the fluctuations to average out, and a constant free energy is considered a trustworthy indication of equilibrium. The free energy stored on average in a subchain is calculated according to:

$$\frac{E}{kT} = \frac{3}{2} \left\langle n_i x_i^2 \cdot e(x_i) \right\rangle \tag{8}$$

$$e(x) = \frac{1}{2} \ln\left(1 - x^2\right) \approx 1 + \frac{x^2}{2} + \frac{x^4}{3} + \frac{x^6}{4} \tag{9}$$

The brackets indicate averaging over all subchains. The function e(x) allows us to take the finite extensibility of chains into account. As with f(x) defined in eq. (5), e(x) is set to one for linear Gaussian chains.

Final results are calculated by averaging at three consecutive levels: **(i)** A first average is computed over all subchains at a given time for a given network realization. **(ii)** This value is averaged over time (under equilibrium conditions). **(iii)** A general average over all realizations is finally calculated, since each network realization has a quenched topology. All final quantities are obtained by equilibrating with several $\Delta t$ values, and extrapolating to $\Delta t = 0$ (e.g., see Honerkamp 1994).

**II.4 Deformation and Stress Tensor**

The deformation is imposed by first moving all beads and the box (and their periodic images) affinely and subsequently relaxing the network. Due to the periodic boundary conditions, after relaxation the network behaves as if the deformation were imposed at infinity.

*Uniaxial extension and simple shear:* All the beads (with Cartesian coordinates x,y,z) are moved according to the following volume preserving laws:

$$\left.\begin{array}{l} x' = x \cdot \lambda \\ y' = y/\sqrt{\lambda} \\ z' = z/\sqrt{\lambda} \end{array}\right\} \text{ uniaxial strain} \qquad (10) \qquad \left.\begin{array}{l} x' = x + \gamma y \\ y' = y \\ z' = z \end{array}\right\} \text{ shear} \qquad (11)$$

where $\lambda$ is the relative elongation and $\gamma$ is the shear strain. The linear dimensions of the initially cubic simulation box $L_x, L_y$ and $L_z$ are also changed according to the above equations. As the nodes in neighbouring simulation boxes are periodic copies of those in the central box, the boxes have to be shifted themselves to assure continuity at the boundary surfaces.

After each multiplicative step in $\lambda$ of magnitude 1.10 (or additive in shear with $\Delta\gamma = 0.10$), the system is equilibrated. The thermodynamic quantity of interest, in our case the stress tensor **T,** is calculated according to the following equation:

$$\mathbf{T} = \upsilon \left\langle \mathbf{F\,a} \right\rangle \qquad (12)$$

In section III, uniaxial extension results are reported in terms of the stress difference $\sigma = T_{xx} - T_{yy}$ generated by the deformation defined in eq. (10). For shear simulations – following equation (11) – the relevant components of the stress tensor are the shear stress $T_{xy}$ and the first and second normal stress differences, respectively, $N_1 = T_{xx} - T_{yy}$ and $N_2 = T_{yy} - T_{zz}$ .

**II.5 Simulation parameters**

All quantities are expressed in units of kT. The Kuhn length is set to b = 0.10, and 50000 beads are located in a simulation box of volume $V = (6.30)^3$, giving a density of beads of $\rho = 200$. This rather high value of the density of beads was taken to obtain a minimal distortion during network formation. Because of the absence of excluded volume interactions in our simulation, the exact numerical value of the density plays no role. We have checked that much lower densities ($\rho = 50$) give equivalent results. It was also verified that the influence of the size of the simulation box on the results is negligible (box volumes from $(5.0)^3$ to $(12.6)^3$). All runs have been done with Gaussian and non Gaussian chains (finite extensibility, eqs.(4) and (5)), for two average numbers of monomers per subchain ($n_o = 50$ and $n_o = 100$, respectively) to check the influence of this parameter. Averages are calculated over at least 10 network realizations. The values of the parameters ($\Theta_b$, $\ell$) of the initial biased random walk are reported in Appendix A. Finally, errorbars are calculated according to standard procedures.

The extrapolation to $\Delta t = 0$ is done by calculating all quantities at three time steps $\Delta t / \tau_R = 0.12, 0.06$, and $0.03$, where $\tau_R = n_o \, b^2 / 6D$ is the Rouse-time of a single subchain (e.g., $\tau_R = 0.166$ with $n_o = 100$). D has been set to unity as only the product $D\Delta t$ enters the calculation. Due to a common numerical problem in Brownian dynamics, the algorithm may adjust $\Delta t$ temporarily to a lower value, and execute the equation of motion for the same total time step. In our case, the problem occurs when chain sliding happens to lead to subchains having very few Kuhn segments. As the modulus of the Brownian force is fixed (and the direction is random) in our implementation, an iteration with a too big time step $\Delta t$ might overstretch the small subchains, leading in the next step to an overcompensation, and so on up to a numerical "explosion" of the system. The frequency of such (avoided) explosions is continually monitored. It depends strongly on the time step $\Delta t$, because the relative importance in eq. (6) of the stochastic force with respect to the spring force scales with $\Delta t^{-1/2}$. The highest value of $\Delta t$ used in this simulation is chosen such that the events of explosions are still extremely rare. Their relative occurrence is less than $10^{-5}$.

## III. SIMULATION RESULTS

### III.1 Equilibrium characteristics

Once a network is constructed we check the result of the connection algorithm. A typical example of results - averaged over 20 realisations - is shown in Table 1. On the left hand side the overall connectivity is displayed, with an average functionality per bead <f> close to 4, implying a high fraction of completely connected beads (f = 4). In the middle of Table 1, the same statistics is shown for endbeads only. The crosslinks formed by them are also in general tetrafunctional. We also verified that the equilibrated network is isotropic. Indeed, the $2^{nd}$ and $4^{th}$ order tensors <**uu**> and <**uuuu**>, where **u** is a unit vector representing the direction of a subchain between two nodes, are very close to their theoretical values for isotropy. On the scale of the chain, isotropy has been verified through **<RR>**.

As outlined in section II, we have used biased random walks to generate the chain conformations before connection in such a way that the chain statistics is Gaussian after connection and equilibration. In the simulation this property corresponds to fulfillment of eqs. (2). The latter, however, are in general not verified for arbitrary network parameters ($\rho$, $\ell$, $n_o$). An example of a mismatch is shown in Fig. 2a, where a purely random walk ("no bias") is shown to lead to a chain where the eqs. (2) are not simultaneously satisfied: the first equation is fulfilled at ca. $n_o = 92$, the second one only at $n_o = 63$ ($N_s = 9$ in this example). The mismatch is due to the fact that $<a^2>$ and $<R^2>/N_s$, though identical immediately after chain formation through a random walk, become different during network formation and equilibration. This is understandable because fluctuations are more effective on the scale of a subchain then on the scale of a chain. The use of an initially biased random walk, following an algorithm presented in Appendix A, allows us to fulfill eqs. (2) after equilibration. Indeed, using a bias parameter of $\Theta_b = 2.43$ shifts up $<R^2>$ much more than $<a^2>$, thereby verifying both eqs. (2) within the errorbars at $n_o = 100$ (cf. the arrow in Fig. 2b).

### III.2 Networks of Gaussian chains

*Uniaxial extension*

We have performed a series of simulations of uniaxial extension (up to $\lambda \approx 10$) for a rubber network of monodisperse Gaussian chains - f(x) is set to 1 in eq. (4) - with various numbers of entanglements per chain: $N_s - 1 = Z - 2 = 0,1,2,3,4,6,8$ and 18. We present two data sets, one with $n_o = 50$ and the other with $n_o = 100$. The resulting stress-strain isotherms are reported in the form of a Mooney plot, $\sigma/(\lambda^2 - 1/\lambda)\nu kT$ vs $1/\lambda$, in Figs. 3a and 3b. At fixed $N_s$, it is found that the Mooney stress decreases with increasing strain. This 'strain softening' must be due to sliplinks, because it does not exist in the completely crosslinked network. Moreover, the stress decreases continuously as $N_s$ increases. A direct consequence is that the modulus $G/\nu kT$ - given by the limit of the Mooney stress as $\lambda \to 1$ - depends on the number of entanglements. It decreases from about ½ (for $N_s = 1$) to a bit more than a third at $N_s = 19$. Notice also that there is no upturn at high strain, which is in line with the Gaussian spring force eq. (4), with f(x) = 1 used in this calculation. As is well known, the theory by James and Guth (1947) for a tetrafunctional network of phantom chains without entanglements gives a horizontal line at $\sigma/(\lambda^2 - 1/\lambda)\nu kT = ½$. Indeed, we obtain this result running our simulation for $N_s = 1$, within errorbars. As can be seen from Figs. 3a and 3b, the number of monomers per subchain $n_o$ has only a marginal influence on the stress-strain isotherms of Gaussian chains.

The comparison with the sliplink network theory by Ball et al (1981) has been presented in our letter (Oberdisse et al 2002). The strain softening at higher strain is substantially less pronounced with our simulation, whereas the agreement in the small-strain limit is very good. As discussed in our previous letter, the reason for the discrepancy at large deformations is in the different excursion of the sliplinks along the chains, which is fixed to the distance between consecutive entanglements in Ball et al (1981), independently of deformation, while it turns out to decrease with increasing deformation in our case. In Fig. 4 the shear modulus $G/\nu kT$ is plotted as a function of the fraction of crosslinks $\phi_c = 1/N_s$, i.e. for Gaussian chains of varying length ($N_s = 1$ to 19), for two numbers of monomers per subchain: $n_o = 50$ and 100. The solid line is the prediction of the network theory by Ball et al (1981). Although it deviates slightly from our data, the observed linearity in $\phi_c$ is remarkable. Here we must keep in mind that the linear relationship between $G$ and $\phi_c$ is an *ingredient* of the theory by Ball et al, and that the limit of $G$ at $\phi_c = 0$ is a function of their (a priori free) sliding parameter $\eta$. In the Figure we have used the value recommended by Ball et al ($\eta = 0.234$). We retain from this discussion that our simulation *predicts* a linear relationship, and that this low-deformation mechanical property is almost consistent with the theory by Ball et al. We also note that the agreement with Ball et al seems better in the case of $n_o = 50$ than for $n_o = 100$.

*Simple shear*

We have performed a series of simulations of simple shear up to deformations of $\gamma = 3.0$ for the same set of entanglements per chain as before, both for $n_o = 50$ and $n_o = 100$. The results for the shear stress $T_{xy}$ are shown in Figs. 5a and 5b, as well as the prediction of the phantom network theory by James and Guth (1947): $\sigma = \frac{1}{2} \gamma kT$ (solid line). Again the non-entangled case ($N_s = 1$) compares favorably with this prediction, whereas the more entangled networks exhibit lower shear stress. The observed decrease in $\sigma$ is also in line with the decrease of the shear modulus $G$ revealed by the Mooney plot (Fig. 4), because $\sigma = G \gamma$ for $\gamma \ll 1$. The first and the second normal stress differences in shear $N_1$ and $N_2$ are reported in Figs. 5c-f. As with the shear stress, increasing the number of entanglements diminishes $N_1$. Note that the Lodge-Meissner relationship holds for all $N_s$ values. The second normal stress difference is zero for complete crosslinked networks ($N_s = 1$).

With increasing number of entanglements ($N_s$ −1), the second normal stress difference $N_2$ (of negative sign) increases in magnitude, but stays weak.

In Figs. 5c and 5d, the phantom network prediction for the first normal stress difference $N_1/\nu kT$ of ½ $\gamma^2$ is confronted with the result for $N_s$ = 1, and again a close agreement is found. The prediction for $N_2$ is simply zero, i.e. no second normal stress difference is expected in the unentangled state, and this is effectively observed in Figs. 5e and 5f.

### III.3 Networks of chains with finite extensibility

*Uniaxial extension*

In analogy with the results of the preceding section, we have performed a series of simulations with networks of chains with finite extensibility, i.e. which obey a force law defined by eqs.(4) and (5). The results for uniaxial extension (up to $\lambda \approx 10$) for monodisperse rubber with various numbers of entanglements per chain: $N_s$ -1 = 0,1,2,3,4,6,8 and 18 have again been obtained for two separate data sets, $n_o$ = 50 and $n_o$ = 100, respectively. The stress-strain isotherms are reported in the form of a Mooney plot, $\sigma/(\lambda^2\text{-}1/\lambda)\nu kT$ vs $1/\lambda$, in Figs. 6a and 6b. At fixed $N_s$, the Mooney stress is seen to first decrease with increasing strain $\lambda$, which is the same strain softening due to the entanglements observed in Gaussian networks. At high strain, however, the stress increases again. This upturn is due to the finite extensibility of the chains. The position of the minimum shifts systematically to higher elongations as $N_s$ increases, and its limiting values as $\phi_c$ tends towards zero, $1/\lambda$ = 0.1615 and $1/\lambda$ = 0.125, for $n_o$ = 50 and $n_o$ = 100, respectively, stay above the totally stretched isolated subchain limit of $1/\sqrt{n_o}$.

A second observation is that the Mooney stress decreases continuously as the number of entanglements per chain $N_s$-1 increases. As a consequence, the modulus G decreases also with increasing $N_s$, which was already shown in our previous letter (Oberdisse et al 2002). Note also that G is higher than in the Gaussian case for the same number of entanglements per chain.

In absence of entanglements, we can check the quality of the simulation by comparing its results to the theory for networks of chains with finite extensibility, i.e. the network model by Edwards and Vilgis (1988). For completely crosslinked networks, they predict a numerical prefactor of the shear modulus of 0.543 ($n_o$ = 50) and 0.521 ($n_o$ = 100), which is quite close to our results (0.532 and 0.513 for $n_o = 50$ and $n_o = 100$, resp.). When comparing these numbers, one must keep in mind that the model needs two parameters, the Ball sliding parameter $\eta$ and the extensibility parameter $\alpha$, whereas there are no free parameters in the simulation. Moreover, there is an ongoing discussion in the literature about the values to be used (Meissner and Matejka 2002), and we have taken the initially proposed parameters (namely, $\alpha = 1/\sqrt{n_o}$; $\eta = 0.234$) by Edwards and Vilgis (1988). In the light of these facts, the agreement is indeed satisfactory. The prediction by Edwards and Vilgis is also superimposed to our data in Fig. 6. For a satisfactory fit, the extensibility parameter $\alpha$ has to be adjusted to lower values, to approximately ½ of $1/\sqrt{n_o}$, which is in line with the qualitative observation discussed before. We interpret this as the ability of the network to redistribute stress from highly strained chains to less strained neighboring chains, a feature which can not be caught by the extensional limit of an isolated chain. As discussed in our letter, and as it was the case for Gaussian networks, the simulated moduli agree nicely with the theoretical predictions for entangled network, whereas there is not enough strain softening at high deformations in the simulation.

*Simple shear*

A series of simulations of simple shear up to deformations of $\gamma = 3.0$ has been performed for the same set of entanglements per chain as before, again both for $n_o = 50$ and $n_o = 100$. We start with the shear stress $T_{xy}$ plotted in Fig. 7a and b. The stress $T_{xy}$ decreases with increasing $N_s$, which corresponds to the diminution of the shear modulus G revealed by the Mooney plot. The first and the second normal stress differences in shear $N_1/\nu kT$ and $N_2/\nu kT$ are reported in Figs. 7c to f. The first normal stress difference $N_1/\nu kT$ decreases with increasing number of entanglements. Note that, unlike in the Gaussian case, the phantom chain prediction deviates considerably from the reported data for $N_s = 1$. The Lodge-Meissner relationship holds for all $N_s$ values. The second normal stress difference $N_2$ increases in magnitude with increasing number of entanglements $N_s - 1$, which proves that $N_2$ is directly related to the entanglements. Surprisingly, $N_2$ is positive in the nonentangled case. Given that we have only changed the force law, and that simulations with the linear force yield a zero second normal stress difference, we think that this is not an artefact of the simulation.

**IV DISCUSSION**

We now analyze the stress tensor predicted by the simulations, focusing on the question of additivity of crosslink and sliplink contributions to the stress tensor, as suggested by the linear dependence of the modulus on the crosslink fraction, Fig. 4. We will see that this allows the determination of the properties of a network of infinitely long chains. Indeed, the contributions from crosslinks and entanglements are often assumed to be additive (Vilgis and Erman 1993) in the theories of rubber elasticity. We follow this idea and temptatively write the total stress as a weighted sum of cross- and sliplink contributions :

$$\frac{\mathbf{T}}{\nu kT} = \varphi_{CL}\mathbf{P} + \varphi_{SL}\mathbf{S} \qquad (13)$$

where $\varphi_{CL} = 1/N_s$ and $\varphi_{SL} = 1 - \varphi_{CL}$ denote the fractions of cross- and sliplinks in the system, respectively. The first member of the rhs of eq. (13), $\mathbf{P}$, represents the classical crosslink contribution, whereas the second one is the contribution of the sliplinks $\mathbf{S}$. In a simulation it would correspond to the stress tensor in the limiting case of a network of infinitely long, endcrosslinked chains ($N_s \rightarrow \infty$). In the case of Gaussian chains, the crosslink contribution has been calculated by James and Guth (1947), and it is given by the Finger tensor $\mathbf{C}^{-1}$ (Macosko 1994):

$$\mathbf{P} = \frac{1}{2}\mathbf{C}^{-1} \qquad (14)$$

For non-Gaussian chains, the simulation result for $N_s = 1$ are taken for $\mathbf{P}$, which correspond to the stress response of a completely crosslinked network.. We then solve eq. (13) for the unknown sliplink contribution, with the stress tensor $\mathbf{T}$, i.e. the outcome of the simulations, as input. The result, $\mathbf{S}$, will in general depend on $N_s$. However, if additivity holds as postulated in eq. (13), then $\mathbf{S}$ will be independent of $N_s$.

*Additivity of stress contributions in uniaxial extension*

We now discuss the sliplink contribution to the stress-strain isotherms of networks of Gaussian and non Gaussian chains together. The remarkable result for $\mathbf{S}$ in uniaxial extension in terms of the relevant components $S_{xx} - S_{yy}$ is shown as a Mooney-Rivlin plot in Fig. 8. In the case of Gaussian chains displayed in the upper graphs, the data for $N_s = 3$ to 19 fall in close vicinity, within approximately 5%, much closer than the pure stress, Fig. 3. Although one can not speak of a mastercurve stricto sensu, this strongly supports the additivity conjectured in eq. (13), at least for large $N_s$. The smallest $N_s$ value ($N_s = 2$, corresponding to one entanglement per chain) shows the strongest deviation, as previously noticed for the modulus (Oberdisse et al 2002). As $N_s$ increases, the curves, although not monotonously approaching, overlap more and more. The result for $N_s = 19$ can be taken as the limiting value of $(S_{xx} - S_{yy})/(\lambda^2 - 1/\lambda)$. A very good fit to this asymptotic curve can be obtained with a quadratic polynomial in $1/\lambda$:

$$\frac{S_{xx} - S_{yy}}{(\lambda^2 - 1/\lambda)} = 0.337 + 0.026(1/\lambda - 1) - 0.116(1/\lambda - 1)^2 \qquad (n_o = 50, \text{ Gaussian}) \qquad (15a)$$

$$\frac{S_{xx} - S_{yy}}{(\lambda^2 - 1/\lambda)} = 0.318 + 0.024(1/\lambda - 1) - 0.102(1/\lambda - 1)^2 \qquad (n_o = 100, \text{ Gaussian}) \qquad (15b)$$

For chains with finite extensibility, shown in the lower graphs of Fig.8, the additivity seems to hold only up to moderate strains. Above $\lambda \approx 3$, the sliplink stresses of the shorter chains ($N_s = 2, 3 \ldots$) deviate systematically. Nonetheless, sliplink contribution of the longest chains takes again a form which can be parametrized as in eq.(15), i.e. the high deformation upturn induced by the crosslinks becomes negligible.

$$\frac{S_{xx} - S_{yy}}{(\lambda^2 - 1/\lambda)} = 0.354 + 0.039(1/\lambda - 1) - 0.098(1/\lambda - 1)^2 \qquad (n_o = 50, \text{ non Gaussian}) \qquad (16a)$$

$$\frac{S_{xx} - S_{yy}}{(\lambda^2 - 1/\lambda)} = 0.326 + 0.036(1/\lambda - 1) - 0.087(1/\lambda - 1)^2 \qquad (n_o = 100, \text{ non Gaussian}) \qquad (16b)$$

All the fitting functions are also shown in Fig. 8. Combining eqs. (13) to (15) for Gaussian chains, e.g., it is now possible to predict the response in uniaxial extension experiments of systems with an arbitrary number of entanglements without any further numerical work. Notice also that the first term in eq. (15) is the entanglement contribution to the shear modulus (normalized to $\nu KT$).

*Additivity of stress contributions in simple shear*

In analogy to the previous section, we have extracted the sliplink contributions to the stress tensor **S** in simple shear. The results for $n_o = 50$ are displayed in Fig. 9 for networks of Gaussian (left-hand side) and non Gaussian (right-hand side) chains, respectively. We start with the shear stress $S_{xy}$, calculated from the simulated stress tensor component $T_{xy}$. As in extension, there is a smaller contribution for short chains ($N_s = 2$), whereas the entanglement contributions of long chains are closer and closer. A good fit to $S_{xy}$ for $N_s = 19$ (which we take again as the limiting curve $N_s \rightarrow \infty$) is given by:

$$S_{xy} = 0.344 \, \gamma - 0.026 \, \gamma^2 \qquad (n_o = 50, \ \text{Gaussian}) \qquad (17a)$$

$$S_{xy} = 0.358 \, \gamma - 0.035 \, \gamma^2 \qquad (n_o = 50, \ \text{non Gaussian}) \qquad (17b)$$

These functions are also shown in Fig. 9. For comparison, we also report the results for the higher number of monomers per subchain ($n_o = 100$):

$$S_{xy} = 0.325 \, \gamma - 0.023 \, \gamma^2 \qquad (n_o = 100, \ \text{Gaussian}) \qquad (18a)$$

$$S_{xy} = 0.333 \, \gamma - 0.023 \, \gamma^2 \qquad (n_o = 100, \ \text{non Gaussian}) \qquad (18b)$$

The first and second normal stress difference can be treated similarly. For the sliplink contribution, parametrization yields the polynomials given in Table 2. For $n_o = 50$, they are also plotted in Fig. 9. As in uniaxial extension, the relevant stress tensor component for systems with arbitrary numbers of entanglements per chain can now be calculated using the additivity, eqs. (13), and the parametrization given in Table 2.

In shear, the comparison with the prediction by Ball et al. (1981) is again favorable for small deformations. As in uniaxial elongation, however, the large strain response in shear is very different from our simulation (Oberdisse et al 2002). For comparison, the prediction for the sliplink contribution to the stress tensor components by Ball et al. is plotted in Figure 9a, c, and e. Equations are given in Appendix B.

Let us now focus on the small-strain properties of **S**. For illustration, in networks of Gaussian chains with $n_o = 100$, the sliplink contributions to the shear modulus determined in uniaxial extension (0.318 $\nu$kT), via the shear stress (0.325 $\nu$kT) or via the first normal stress difference (0.321 $\nu$kT) are reasonably close, indicating an over-all coherence of our simulations. The same is true for $n_o = 50$, where the sliplink contribution is slightly higher. Given that the crosslink contribution to the shear modulus in the phantom chain model is ½ $\nu$kT, we come to one of the main results of our simulation: the sliplink contribution is about ⅔ of the crosslink contribution. Note that this is in good agreement with the value predicted by Ball et al ($1/2(1 + \eta)^2 \approx 0.33$) in their analytical theory (Ball et al 1981, Oberdisse et al 2002).

The same argument holds with chains with finite extensibility. The shear modulus can be determined by uniaxial strain, the shear strain, or the first normal stress difference, with little numerical differences. As with Gaussian chains, it is higher for a lesser number of monomers per subchain. The ratio of the sliplink contribution to the one from the crosslinks is $0.357/0.532 = 0.67$ for $n_o = 50$, and $0.330/0.513 = 0.64$ for $n_o = 100$, i.e. it is again close to ⅔. Note that this corresponds to the results of Edwards and Vilgis, who predict $0.363/0.543 = 0.67$ for $n_o = 50$, and $0.343/0.521 = 0.66$ for $n_o = 100$, using again $\eta = 0.234$ and the extensibility parameter $\alpha = 1/\sqrt{n_o}$. (Edwards and Vilgis 1988).

The magnitude of the normal stress differences also deserves some comments. In agreement with experimental results for polymer *melts*, the second normal stress difference is a lot smaller (and of opposite sign) than the first normal stress difference. More specifically, the following material parameters for networks of infinitely long chains (only slip-links, $N_s \rightarrow \infty$) can been read off in the small deformation limit:

$$A_1 = \lim_{\gamma \rightarrow 0} \frac{S_{xx} - S_{yy}}{\gamma^2} \tag{19a}$$

$$A_2 = -\lim_{\gamma \rightarrow 0} \frac{S_{yy} - S_{zz}}{\gamma^2} \tag{19b}$$

The ratio $A_1/A_2$ of the second to the first normal stress difference at vanishing strain can be calculated from the polynomials given in Table 2. It is found to be $0.115 \pm 0.01$. This is interesting, because the constitutive equations of polymer *melts* differ considerably in the prediction of the ratio $\Psi = - N_2/N_1$ close to mechanical equilibrium. For instance, the tube model by Doi and Edwards predicts $\Psi = 1/7$, and $\Psi = 2/7$ with the independent alignment approximation (Doi and Edwards 1986), while experimental results seem to lie around ¼ (Marrucci et al 2000).

## V CONCLUSION

A primitive chain Brownian dynamics simulation for elastomers with many entanglements has been presented. Our model is a numerical implementation of the sliplink network model for Gaussian chains by Ball et al (1981), and its finite extensibility version by Edwards and Vilgis (1988). Like these theories, our simulation scheme is based on crosslinks (permanent junctions) and sliplinks (sliding junctions representing entanglements), and subchains connecting such nodes. It has been used to determine the mechanical response to uniaxial strain and simple shear of highly entangled networks of monodisperse chains, with up to 18 entanglements per chain.

We have tested the additivity assumption for cross- and sliplinks which is usually used in the analytical literature. Additivity is found for long enough chains (number of subchains $N_s > 3$), not too close to the finite extensibility limit of individual chains. This analysis lead us to the contribution **S** of sliplinks to the stress tensor, and we have determined its relevant tensorial components.

The fact that comparison of our simulations with the theory of Ball et al. is fully satisfactory in the linear range while showing large discrepancies in the nonlinear domain needs some discussion. One possibility might be the use in our simulations of elastic forces for monomer motion across sliplinks rather than chemical potential gradients. The chemical potential $\mu$ is readily derived from the classical expression for the free energy $E$ of one subchain, cf. eq. (8):

$$\mu = -\frac{\partial E}{\partial n} = \frac{3}{2}kT\frac{a^2}{n^2 b^2} \tag{20}$$

and therefore the gradient along the chain can be written as

$$grad\ \mu = \frac{\mu_j - \mu_i}{\frac{1}{2}(a_j + a_i)} = 3\frac{kT}{b^2}\frac{a_j^2/n_j^2 - a_i^2/n_i^2}{a_j + a_i} \tag{21}$$

The last expression could replace the elastic force difference used in Eq. (7a), i.e., the difference (in the Gaussian range) $(3kT/b)(x_j - x_i)$. We have not tried this variant in the course of this work, but we must admit that the treatment of Ball et al., being based on free energy, should automatically account for chemical potential gradients.

Still in the nonlinear range, the observed discrepancy of our simulations with respect to actual data, can perhaps be explained also by a fundamental inadequacy of the model in the way entanglement are represented. We here refer to the fact that entanglements are depicted (here as well as in the theory of Ball et al.) as sliplinks between chain pairs. In reality the topological obstacles are not holonomic in nature, i.e., the uncrossability constraint between interacting chains is active only in one direction, and is no longer active when chains move one away from the other. This aspect is not captured in sliplink models. Of course, for this aspect to be relevant in explaining the discrepancy, it is required that the non-holonomic nature of the interaction becomes more important with increasing deformation.

**Acknowledgements:** Work primarily supported by the EU under contract number FMRX–CT98-0210. Parts of this work were conducted within the scientific programme of the Network of Excellence 'Soft Matter Composites: an approach to nanoscale functional materials' supported by the European Commission.

**APPENDIX A: Bias-algorithm for the initial chain conformation**

Network construction (chain linking) and subsequent equilibration changes the conformation of the chains. In order to obtain a Gaussian statistics according to eqs. (2) after equilibration we bias the random walk by imposing an orientational coupling between neighboring subchains i and i+1:

$$\mathbf{a}_i \cdot \mathbf{a}_{i+1} / (a_i a_{i+1}) = \cos \Theta_i \tag{A1}$$

These angles $-\pi < \Theta_i < \pi$ are drawn from a (non normalized) distribution function:

$$P(\Theta_i) = \exp(-\Theta_i^2/2\Theta_b^2) \quad \text{for} \quad -\pi < \Theta_i < \pi$$

$$= 0 \quad \text{elsewhere} \tag{A2}$$

The azimuthal angle is drawn from a uniform distribution between 0 and $2\pi$. The additional parameter $\Theta_b$ characterizes the width of the distribution. The smaller the width, the more swollen the chain. Note that the chain is not stretched in the sense that there is no preferred orientation.

For a chain of Z beads, we can now freely choose $n_o$ and the subchain density $\nu$, and verify eqs. (2) by choosing the initial steplength of the biased random walk $\ell$ and the swelling parameter $\Theta_b$. The only limit on $\nu$ is that a too small density might make the connection difficult, leading to too large distortions of the chain. As a rule of thumb, the density must be chosen such that the search radii given in eqs. (1) are smaller, say by a factor 3, than the subchain length, fixed through $n_o$.

The parameters used in all our simulations with $\rho = 200$ are given in Table 3. In general, eqs. (2) for the Gaussian statistics after equilibration are verified within less than a percent, e.g. for $n_o = 100$, 20 realizations, $N_s = 7$ in 6250 chains: $<a_i^2> = 1.000 \pm 0.002$, $<R^2>/N_s = 1.002 \pm 0.005$.

**APPENDIX B: Shear stress tensor prediction for sliplinks by Ball et al**

We have calculated the sliplink contribution to the components of the stress tensor under simple shear from the expression by Ball et al. (1981):

$$
\begin{aligned}
S_{xy} &= \frac{\gamma}{2} \frac{(1+\eta)^2 + \eta^2 \gamma^2}{\left[(1+\eta)^2 + \eta \gamma^2\right]^2} \\
S_{xx} - S_{yy} &= \gamma \, S_{xy} \\
S_{yy} - S_{zz} &= -\frac{\eta \gamma^2}{2} \frac{2(1+\eta) + \eta \gamma^2}{\left[(1+\eta)^2 + \eta \gamma^2\right]^2}
\end{aligned}
\tag{B1}
$$

**Table Captions:**

Table 1:      Network connection and statistics of equilibrated network of Gaussian chains ($N_s =$ 9, $N_c = 5000$, $n_o = 100$, $\rho = 200$, 20 realizations).

Table 2:      Polynomials describing the sliplink contributions to the stress tensor.

Table 3:      Parameters used to generate biased random walks. $n_o$ denotes the average number of monomers per subchain, $N_s = Z - 1$ the number of subchains per chain, $\ell$ the initial steplength of the random walk generating the chains, and $\Theta_b$ the bias parameter used in eq. (A2). $\Theta_b = \infty$ is the limiting case of a true random walk.

**Figure Captions:**

Figure 1:     Illustration of the two types of connections between chains: A sliplink connects two chains allowing monomer transport along each chain, whereas a crosslink connects up to four chain ends.

Figure 2:     Plot of the average square subchain length $<a^2>$, the rescaled average square chain length $<R^2>/N_s$, and their theoretical value $n_o b^2$ as a function of $n_o$ for (2a) an inappropriate bias parameter $\Theta_b$, (2b) an appropriate bias parameter ($\Theta_b = 2.43$). The triangular area in (2a) illustrates the difficulties in fulfilling the Gaussian statistics. ($N_s = Z - 1 = 9$, $N_c = 5000$, $\rho = 200$, 10 realizations).

Figure 3:     Mooney plot of the stress-strain isotherm of a network of Gaussian chains with (3a) $n_o = 50$ and (3b) $n_o = 100$. Symbols are simulation results for different chain lengths, $N_s = 1$ to $N_s = 19$ subchains per chain.

Figure 4:     Shear modulus $G/\nu kT$ as a function of crosslink fraction $\phi_c = 1/N_s$ for a network of Gaussian chains ($n_o = 50$ (●) and 100 (○), $N_c = 5000$). The continuous line is the prediction by Ball et al, the broken line a linear curve fit for $n_o = 100$.

Figure 5:  Simulation results of (5a,5b) the shear stress $T_{xy}$, (5c,5d) the first normal stress difference $N_1$ and (5e,5f) the second normal stress difference $N_2$ as a function of deformation $\gamma$ for a network of Gaussian chains ($n_o = 50$ on the left and $n_o = 100$ on the right). The solid line is the prediction of the phantom chain model for the completely crosslinked network.

Figure 6:  Mooney plot of the stress-strain isotherm of a network of non Gaussian chains with (6a) $n_o = 50$ and (6b) $n_o = 100$. Symbols are simulation results for different chain lengths, $N_s = 1$ to $N_s = 19$ subchains per chain. The solid line is the prediction by Edwards and Vilgis for completely crosslinked networks, with extensibility parameter $\alpha = 0.07$ ($n_o = 50$) and $\alpha = 0.05$ ($n_o = 100$).

Figure 7:  Simulation results of shear stress $T_{xy}$, the first normal stress difference $N_1$ and the second normal stress difference $N_2$ as a function of deformation $\gamma$ for a network of non Gaussian chains ($n_o = 50$ on the left and $n_o = 100$ on the right).

Figure 8:  Mooney plot of the sliplink contribution to the stress-strain isotherm extracted according to eq. (13) for networks of (8a,8b) Gaussian and (8c,8d) non Gaussian chains. The solid line is a fit to the data at the highest $N_s$ value. Symbols: ($\circ$) $N_s = 2$, ($\bullet$) $N_s = 3$, ($\square$) $N_s = 4$, ($\blacksquare$) $N_s = 5$, ($\diamond$) $N_s = 7$, ($\blacklozenge$) $N_s = 9$, ($\times$) $N_s = 19$.

Figure 9:  Sliplink contribution to the stress tensor of networks of Gaussian (on the left) and non Gaussian (on the right) chains to (9a,9b) the shear stress, (9c,9d) the first and (9e,9f) the second normal stress difference. Only data for $n_o = 50$ is shown. The solid line is a fit to the data at the highest $N_s$ value, the dotted one is the prediction by Ball et al. The error bars are not shown for clarity; they are of the same order of magnitude as in the other Figures. Symbols: ($\circ$) $N_s = 2$, ($\bullet$) $N_s = 3$, ($\square$) $N_s = 4$, ($\blacksquare$) $N_s = 5$, ($\diamond$) $N_s = 7$, ($\blacklozenge$) $N_s = 9$, ($\times$) $N_s = 19$.

**TABLES:**

| All beads : | | Endbeads only : | | Tensorial order parameter S = <uu> | | |
|---|---|---|---|---|---|---|
| Average functionality | < f > =3.97 | Average functionality | < f >= 3.93 | 1/3 ± 0.0001 | 0.000 ± 0.001 | 0.000 ± 0.001 |
| f=1: | 0.0034% | f=1: | 0.017% | | 1/3 ± 0.0001 | 0.000 ± 0.001 |
| f=2: | 1.19% | f=2: | 1.23% | | | 1/3 ± 0.0001 |
| f=3: | 0.84% | f=3: | 4.22% | **Higher moments :** | | |
| f=4: | 97.96% | f=4: | 94.53% | $<x^4>=<y^4>=<z^4>=1/5 \pm 0.0002$ | | |
| Fraction of unconnected beads: | 0.95% | Bead fraction with f=1 or f=2: | 1.25% | $<x^2y^2>=<x^2z^2>=<y^2z^2>=1/15 \pm 0.001$ <odd terms > = 0 ± 0.001 | | |

Table 1
(Oberdisse et al)

|  | $n_o = 50$ | $n_o = 100$ |
|---|---|---|
| Gaussian | $S_{xx} - S_{yy} = 0.342 \, \gamma^2 - 0.025 \, \gamma^3$<br><br>$S_{yy} - S_{zz} = -0.0415 \, \gamma^2 + 0.0074 \, \gamma^3$ | $S_{xx} - S_{yy} = 0.321 \, \gamma^2 - 0.021 \, \gamma^3$<br><br>$S_{yy} - S_{zz} = -0.0340 \, \gamma^2 + 0.0063 \, \gamma^3$ |
| Non Gaussian | $S_{xx} - S_{yy} = 0.358 \, \gamma^2 - 0.025 \, \gamma^3$<br><br>$S_{yy} - S_{zz} = -0.0441 \, \gamma^2 + 0.0077 \, \gamma^3$ | $S_{xx} - S_{yy} = 0.332 \, \gamma^2 - 0.022 \, \gamma^3$<br><br>$S_{yy} - S_{zz} = -0.0389 \, \gamma^2 + 0.0067 \, \gamma^3$ |

Table 2
(Oberdisse et al)

| Chains with finite extensibility: | | | | | |
| --- | --- | --- | --- | --- | --- |
| $n_o = 50$ | | | $n_o = 100$ | | |
| $N_s$ | $\ell$ | $\Theta_b$ | $N_s$ | $\ell$ | $\Theta_b$ |
| 1 | 0.652 | $\infty$ | 1 | 0.962 | $\infty$ |
| 2 | 0.595 | 1.95 | 2 | 0.885 | 2.20 |
| 3 | 0.595 | 2.25 | 3 | 0.87 | 2.35 |
| 4 | 0.595 | 2.30 | 4 | 0.87 | 2.45 |
| 5 | 0.595 | 2.35 | 5 | 0.87 | 2.53 |
| 7 | 0.600 | 2.45 | 7 | 0.87 | 2.60 |
| 9 | 0.600 | 2.50 | 9 | 0.87 | 2.65 |
| 19 | 0.605 | 2.55 | 19 | 0.87 | 2.65 |
| **Gaussian chains (infinite extensibility):** | | | | | |
| $n_o = 50$ | | | $n_o = 100$ | | |
| 1 | 0.64 | $\infty$ | 1 | 0.96 | $\infty$ |
| 2 | 0.58 | 1.77 | 2 | 0.87 | 2.0 |
| 3 | 0.58 | 2.03 | 3 | 0.858 | 2.15 |
| 4 | 0.58 | 2.10 | 4 | 0.855 | 2.25 |
| 5 | 0.58 | 2.15 | 5 | 0.857 | 2.38 |
| 7 | 0.585 | 2.25 | 7 | 0.858 | 2.45 |
| 9 | 0.585 | 2.28 | 9 | 0.856 | 2.43 |
| 19 | 0.587 | 2.30 | 19 | 0.856 | 2.45 |

Table 3
(Oberdisse et al)

**Figures:**

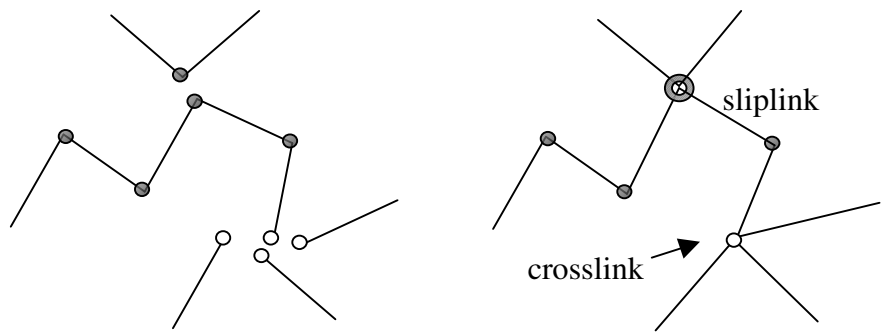

Figure 1
(Oberdisse et al)

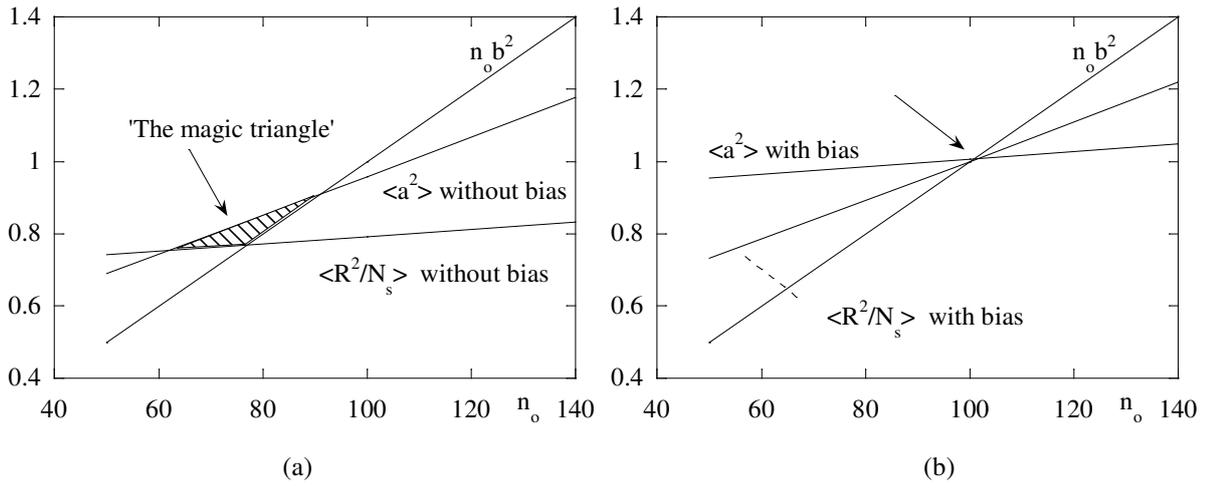

(a)

(b)

Figures 2 a and 2b
(Oberdisse et al)

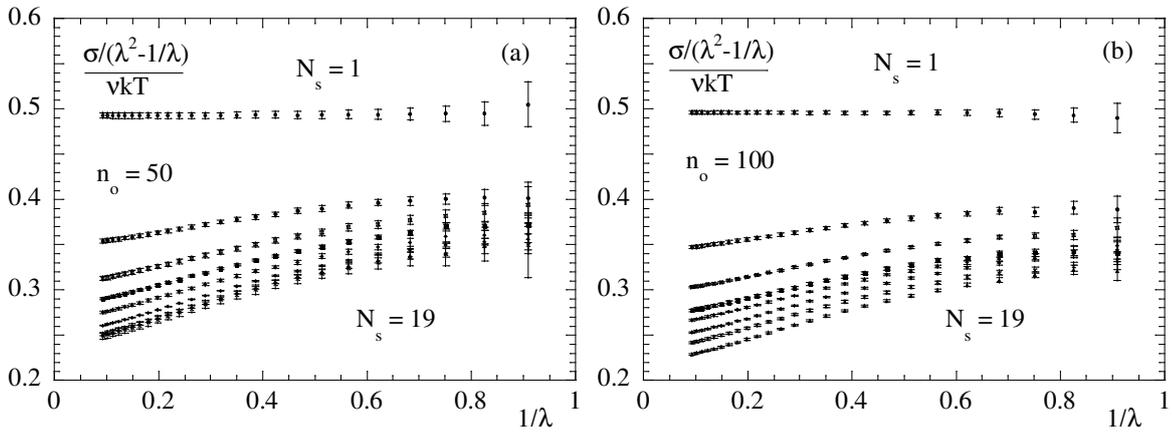

Figure 3
(Oberdisse et al)

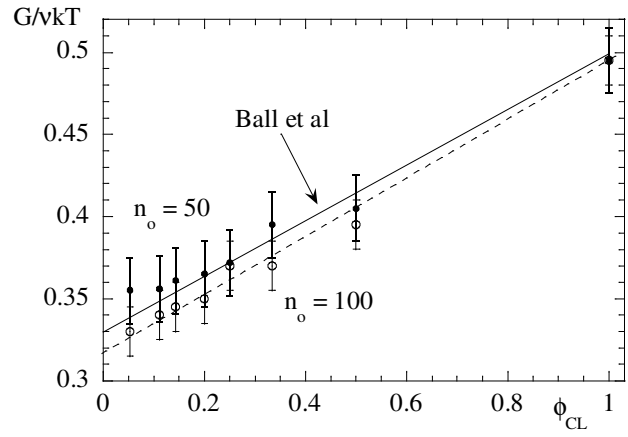

Figure 4
(Oberdisse et al)

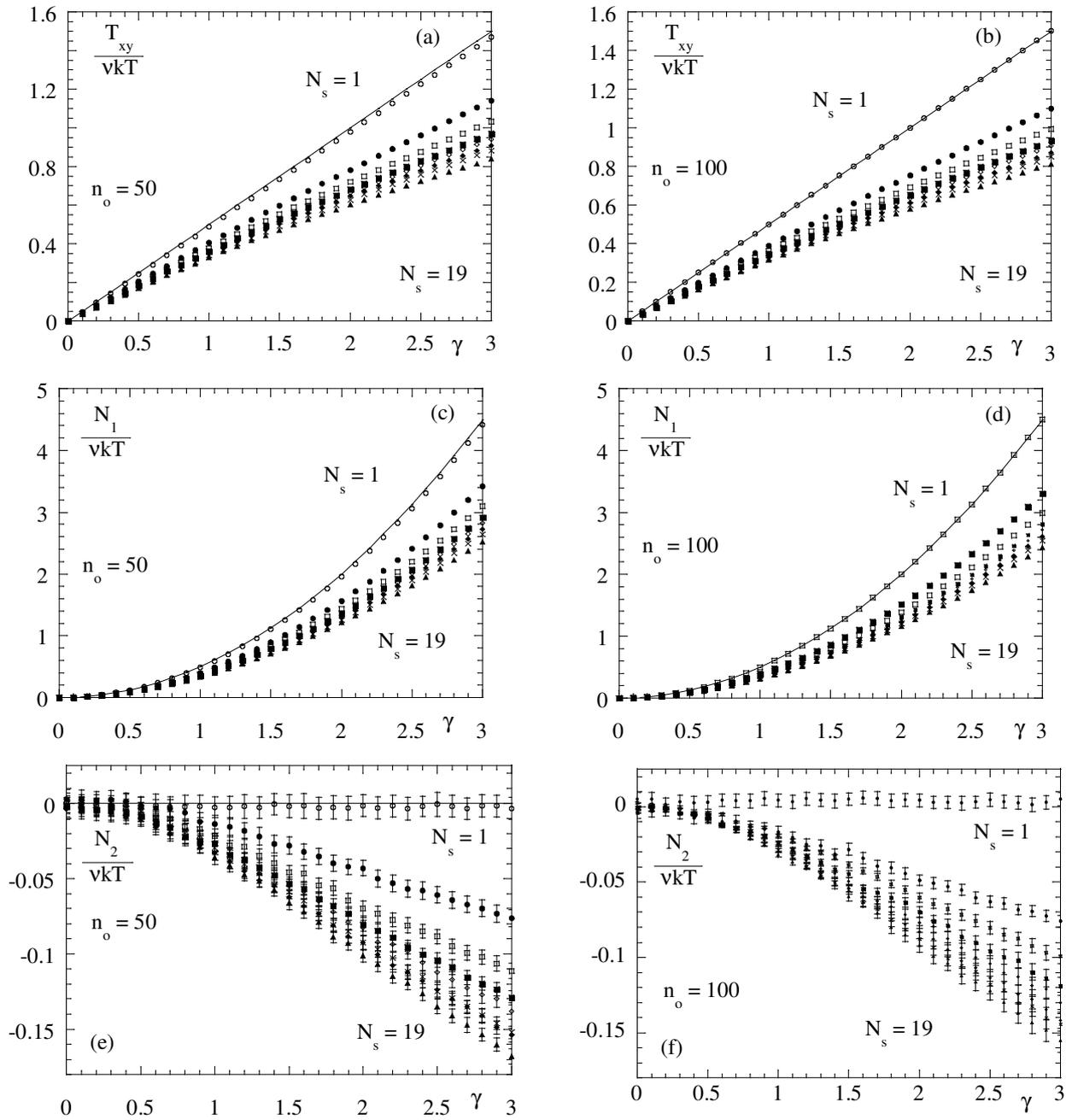

Figure 5
(Oberdisse et al)

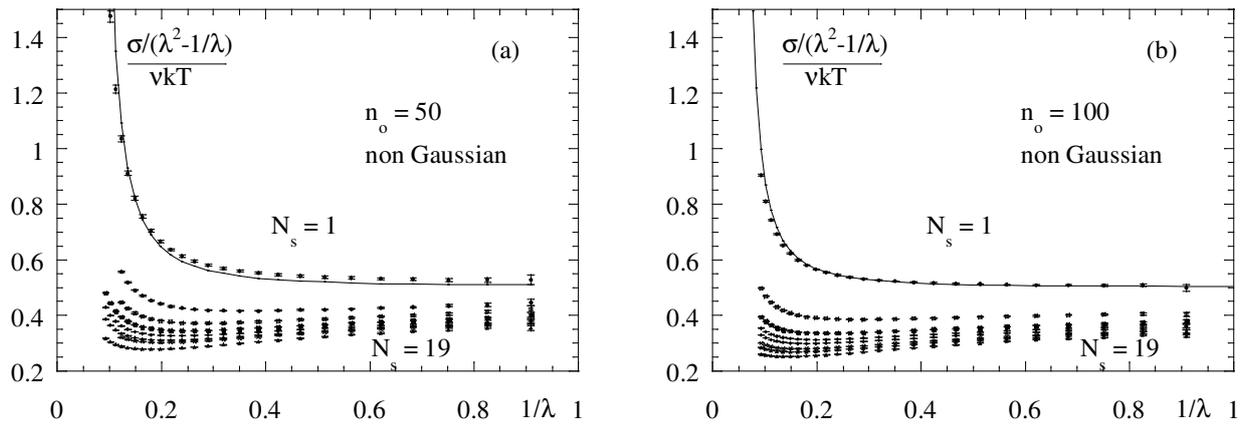

Figure 6
(Oberdisse et al)

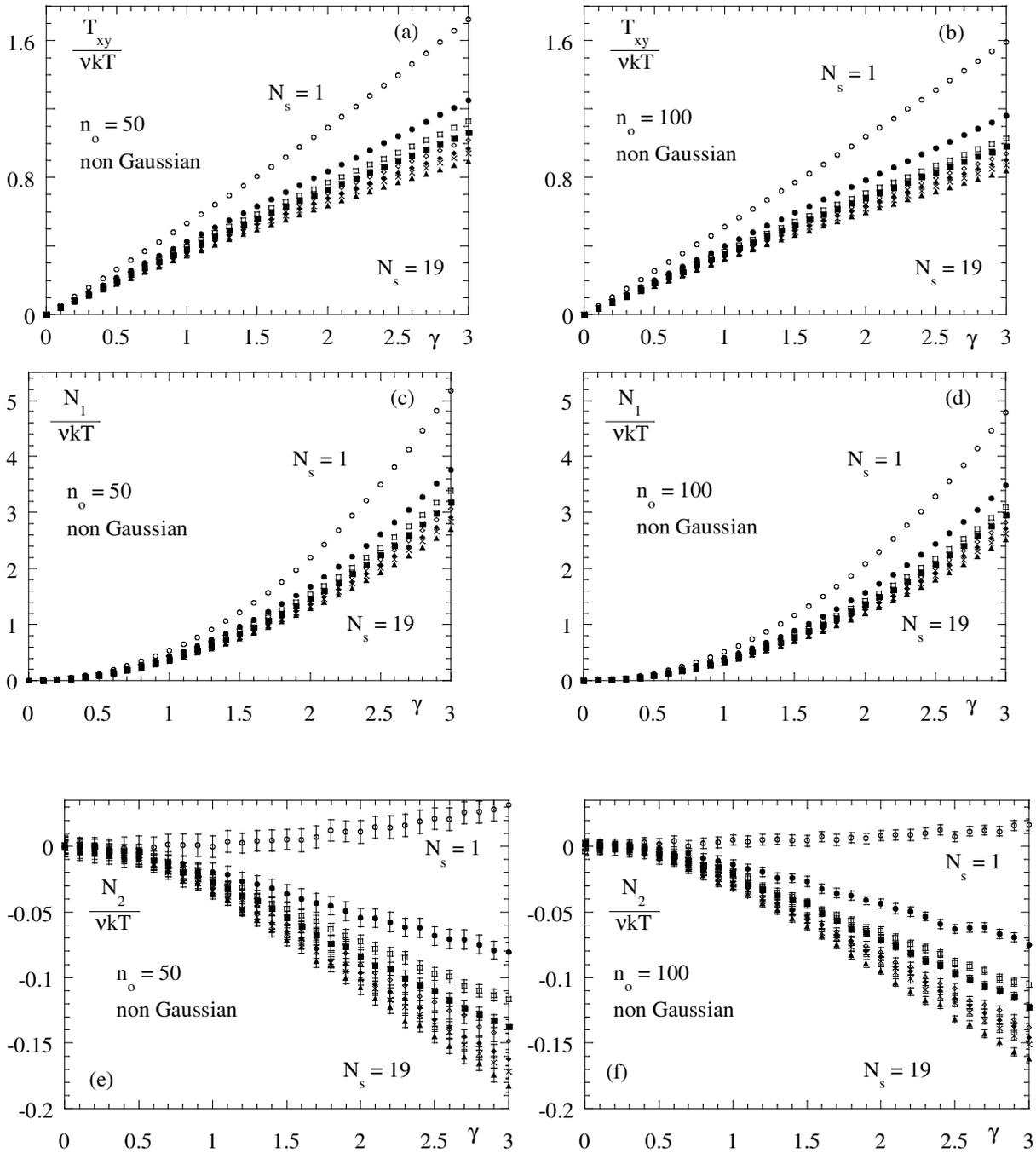

Figure 7
(Oberdisse et al)

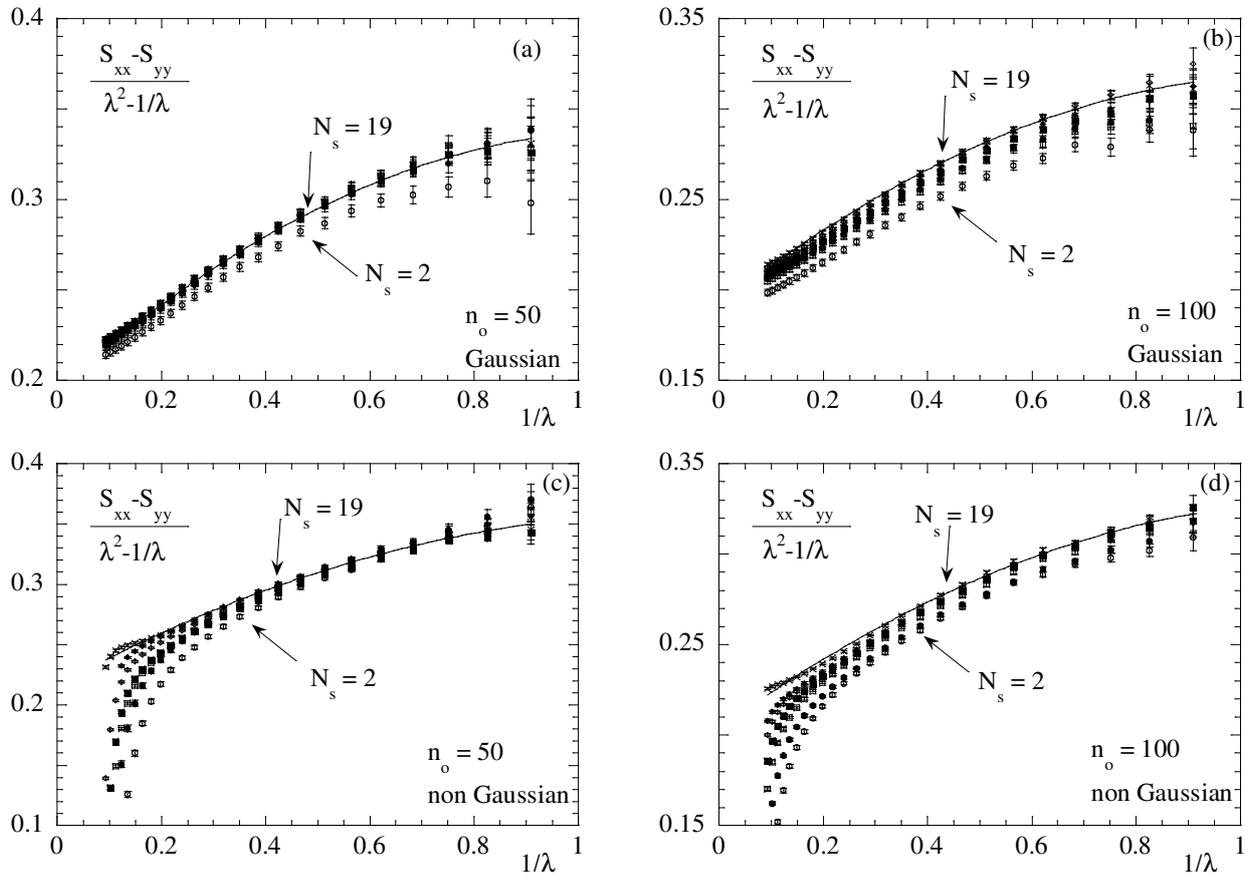

Figure 8
(Oberdisse et al)

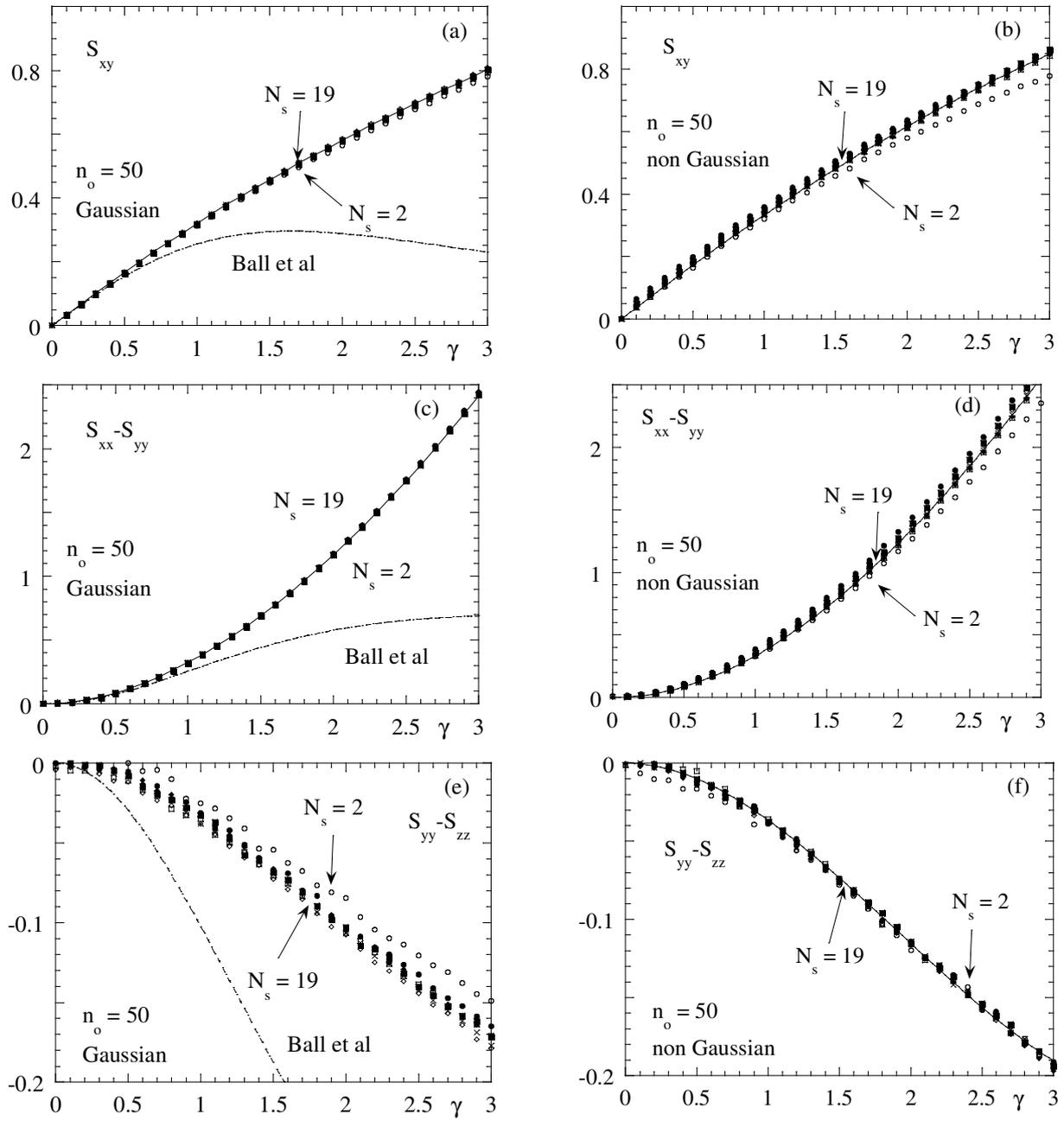

Figure 9
(Oberdisse et al)